# A route to high peak power and energy scaling in the mid-IR chirped-pulse oscillator-amplifier laser systems

**ALEXANDER RUDENKOV,**[1,*] **VLADIMIR L. KALASHNIKOV,**[1]
**EVGENI SOROKIN,**[2,3] **MAKSIM DEMESH,**[1] **AND IRINA T. SOROKINA**[1,3]

[1]*Department of Physics, Norwegian University of Science and Technology, N-7491 Trondheim, Norway*
[2]*Photonics Institute, Vienna University of Technology, 1040 Vienna, Austria*
[3]*ATLA Lasers AS, Richard Birkelands vei 2B, 7034 Trondheim, Norway*
*\* alexander.rudenkov@ntnu.no*

**Abstract:** The paper introduces a new route towards the ultrafast high laser peak power and energy scaling in a hybrid mid-IR chirped pulse oscillator-amplifier (CPO-CPA) system, without sacrificing neither the pulse duration nor energy. The method is based on using a CPO as a seed source allowing the beneficial implementation of a dissipative soliton (DS) energy scaling approach, coupled with a universal CPA technique. The key is avoiding a destructive nonlinearity in the final stages of an amplifier and compressor elements by using a chirped high-fidelity pulse from CPO. Our main intention is to realize this approach in a $Cr^{2+}$:ZnS-based CPO as a source of energy-scalable DSs with well-controllable phase characteristics for a single-pass $Cr^{2+}$:ZnS amplifier. A qualitative comparison of experimental and theoretical results provides a road map for the development and energy scaling of the hybrid CPO-CPA laser systems, without compromising pulse duration. The suggested technique opens up a route towards extremely intense ultra-short pulses and frequency combs from the multi-pass CPO-CPA laser systems that are particularly interesting for real-life applications in the mid-IR spectral range from 1 to 20 μm.

## 1. Introduction

In the last decades, many front-end achievements in modern science have been made possible by the progress in femtosecond laser technology [1, 2]. These advances have provided breakthroughs, for example, in attosecond and relativistic physics [3, 4] and ultrafast spectroscopy [5], nano-photonics [6], particle beam acceleration [7], material processing [8], and many other fields of science and technology. Such ultra-intense ultrashort light pulses opened the doors for the discovery of many new fundamental physical phenomena, allowing us to look inside the atom, deliver the power to objects in outer space, produce ultraprecise measurements, make new discoveries in astrophysics, and quantum electrodynamics, open the doors to relativistic phenomena and even question some of the fundamental constants. The current status of femtosecond technology is PW-level peak power lasers [9-11]. However, even more moderate albeit still very high - in excess of MWs - peak powers, in particular when produced in the mid-IR "molecular fingerprint" wave-length range, open up a wide avenue towards numerous scientific and industrial applications such as fine material processing of semiconductors and composite materials [12]. It is also important for many industrial applications, that the next generation of high-intensity lasers become more compact, user-friendly, and work at high repetition rates (over MHz). These tabletop systems providing sub- and over-μJ femtosecond pulses directly from a mode-locked laser can be readily used for high sensitivity environmental monitoring (e.g. for sensing of hazardous gases in the atmosphere) or for sub-surface processing of silicon [13]. In the future compact high-intensity CPO-CPA lasers based on such seed sources as proposed in this paper will also make numerous societal applications a reality, with proton therapy in medicine, a clean (neutron- and nuclear waste-less) nuclear energy production with particularly compact femtosecond laser drivers, nuclear waste transmutation, just to name a few.

To date, the solid-state mode-locked lasers have allowed the generation of femtosecond high-peak power pulses directly from an oscillator with high (>MHz) repetition rates [14]. In contrast to a classical chirped-pulse amplification scheme [15], such a high pulse repetition rate promises an extreme signal rate improvement factor of $10^3$-$10^4$. However, the main trouble in further energy scaling in parallel with peak power for such systems is still the destructive contribution of nonlinearities. The alternative approach is provided by the integration of a *chirped-pulse amplifier* (CPA) and the laser into a single oscillator, *chirped pulse oscillator* (CPO) [16, 17], which utilizes the remarkable property of a *dissipative soliton* (DS) to combine high stability with the energy scalability [18]. High-energy seed pulse from CPO would allow reducing the number of amplifying cascades, enhance stability, improve high repetition rate dynamics in Regenerative Amplifiers [19], increase temporal intensity contrast in high-power CPA laser systems [20] and prevent excessive noise accumulation. The feasibility of this approach was also demonstrated for the fiber laser-amplifier systems [21, 22].

Simultaneously, the deep insight into the mechanisms of ultrashort pulse generation allows the development of a "smart laser" in which a transparent algorithm of optimization and, ultimately, self-optimization of a laser device would be realized [23, 24]. In particular, the concepts of a *dissipative soliton resonance* [25] and a *master diagram* [26] provide us with the tools for such an optimization aimed at a DS energy scaling (e.g., see [27, 28]).

In our work, we theoretically and experimentally consider the ways of ultrafast laser power and energy scaling in a hybrid mid-IR CPO-CPA system. Our main intention is to use a CPO based on a $Cr^{2+}$:ZnS-based mode-locked laser [29, 30] as a source of energy-scalable DS with well-controllable phase characteristics for a single-pass $Cr^{2+}$:ZnS polycrystalline ceramic amplifier. Using a CPO as a seed source allowed us to implement a DS energy scaling approach and incorporate such laser to the universal concept of CPA, beneficially without using a pulse stretcher and reducing the amplification stage that lowers the system complexity and avoids a destructive nonlinearity contribution. DS energy scaling can be performed by changing several laser parameters such as pump power, mode area, and pulse repetition frequency (cavity period). This issue will be discussed in more detail later in this work.

$Cr^{2+}$:ZnS crystals are characterized by the high stimulated emission cross sections and one of the broadest among all existing laser gain [13], which makes it possible to obtain sufficient amplification of a few optical cycle pulses even in a single-pass amplifier configuration. A combination of more than one amplification stage can provide comparatively high pulse energy still at a full repetition rate, while the nonlinearity in the amplifier media is controlled by scaling the mode area, thus operating the amplifier close to the linear regime. What is particularly important is that using a CPO provides an effective pre-amplification without degradation of amplification efficiency due to operation in the vicinity of the maximum *fidelity* range, where chirp is almost spectrally independent.

In this paper, we propose a novel technique to provide hybrid CPO-CPA pulse energy scaling while not having to sacrifice the pulse duration (maximizing peak power) and preserving the output spectrum's high-fidelity compression. In the following two sections, we describe the first steps in this direction, both experimentally and theoretically, demonstrating the viability of the proposed concept.

## 2. Experimental setup

For the experimental part of our work, we developed a setup consisting of a seed laser, a single-pass amplifier, and a prism compressor (Figure 1). For pulse parameter control we used the home-made interferometric autocorrelator, capable to measure pulse duration up to 50 ps, APE waveScan Extended IR spectrometer (800-2600 nm), and usual sensors like power meters and photodetectors.

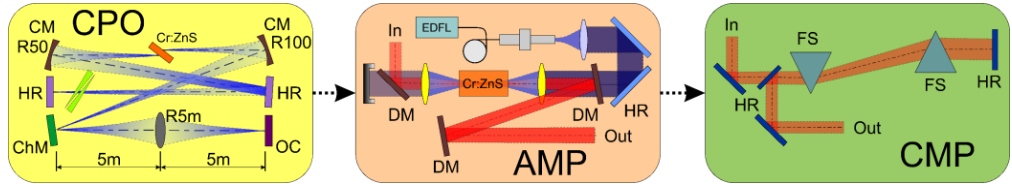

Fig. 1. Experimental setup. Chirped pulse oscillator (CPO), single pass amplifier (AMP), prism compressor (CMP), concave mirror (CM), chirped mirror (ChM), output coupler (OC), dichroic mirrors (DM), erbium doped fiber laser (EDFL), fused silica prisms (FS), highly reflective mirrors (HR).

We start the design of the seed laser from a 70 MHz soliton-like pulse Kerr-lens mode-locked laser operating in anomalous dispersion regime, with 2.6 mm thick $Cr^{2+}$:ZnS Brewster-cut single crystal as an active medium. Beam radius in the active element was about 40 um. As a pump source we employ a 5 W Er-fiber laser operating at 1610 nm. Intracavity dispersion balance was maintained by using specially designed chirped mirrors which compensated simultaneously the second (group delay dispersion, GDD) and the third (third-order dispersion, TOD) orders of dispersion created by gain medium and the YAG wedges, used for fine tuning of the GDD. Highly reflective mirrors and an output coupler were also designed for broadband operation and flat dispersion profile with low absolute values. Such approach allowed obtaining soliton-like pulses with about 36 fs duration, assuming a $sech^2$ shape. Spectra and autocorrelation traces are shown in Figs. 2 and 3. However, direct amplification of such pulses inevitably causes strong nonlinear effects in the amplifier medium [31].

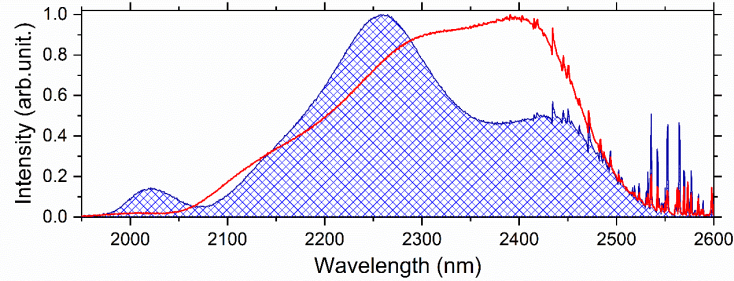

Fig. 2. Soliton-like pulse laser spectra.

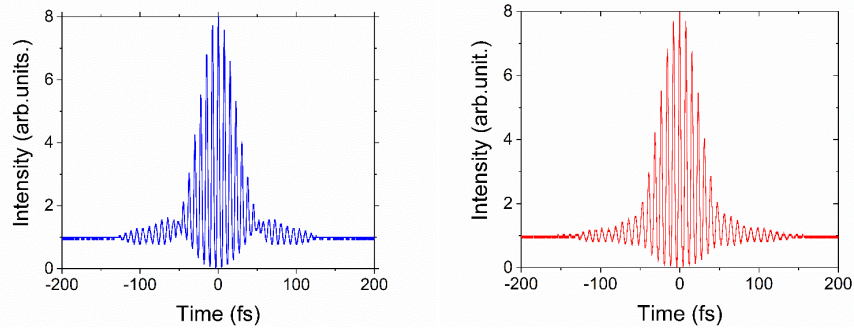

Fig. 3. Interferometric autocorrelation traces of soliton-like pulses.

To switch the laser to CPO that generates dissipative solitons, the cavity was extended using broadband highly reflective mirrors with small positive GDD. CPO sketch in Figure 1 shows a generalized cavity scheme for understanding the principle of its operation. The real experimental setup included the following optical elements: a 2.6 mm thick Brewster-oriented $Cr^{2+}$:ZnS active element, one chirped mirror (1 reflection), HR mirrors of type 1 (10 refl.), HR mirrors of type 2 (8 refl.), YAG wedges, and an output coupler. The corresponding dispersion profiles (in total per round-trip for each group of elements) are shown in Figure 4a together with full cavity round-trip GDD. The "R5m" element, schematically shown as a lens, is a mirror with 5 m radius of curvature. With some folding mirrors (not shown in Fig. 1) the round-trip path became over 24 m. Pulse repetition frequency of such a long cavity was 12.345 MHz that allowed CPO pulse energy scaling. We have also took into account the atmospheric dispersion [32] while calculating the total GDD profile in Figure 4a.

Reflection spectra of mirrors used in CPO are shown in Figure 4b in total for each group of mirrors.

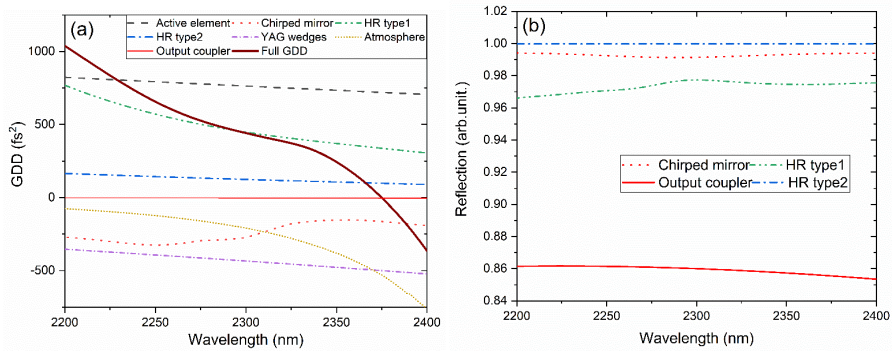

Fig. 4. CPO elements dispersion profiles together with full amount of GDD for cavity roundtrip (a) and mirrors reflection curves (b) (both graphs illustrates GDD and reflection characteristics in total for each group of elements).

Single pass amplifier used a $Cr^{2+}$:ZnS polycrystalline gain element with 13.5 mm length and 8x2.7 mm$^2$ cross-section with $Cr^{2+}$ concentration of $2.3 \cdot 10^{18}$ cm$^{-3}$ (IPG Photonics). CPO beam radius in the amplifier gain element was about 35 μm, the same as the pump one. The amplifier acts as a CW amplifier, because pumping is continuous and $Cr^{2+}$ lifetime of ~5 μs is much longer than the pulse repetition rate of ~80 ns.

After the amplifier we install a prism compressor made of very dry fused silica material. Total dispersion profile of the amplifier and compressor elements is shown in Figure 5.

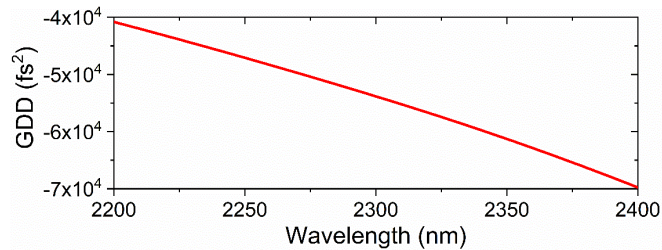

Fig. 5. Combined dispersion of the amplifier and the prism compressor.

## 3. Experimental results

During the experiment we first characterize the laser parameters in continuous wave (CW) regime of operation. Then we switch to the mode-locking (ML) regime. ML operation was provided by a Kerr-lens mechanism only, that is better suitable for energy scaling purposes due to higher laser induced damage threshold compared to using absorbing material modulators such as SESAMs or graphene mirrors [33-35].

Output power characteristics of the laser in both CW and ML regimes are shown in Figure 6.

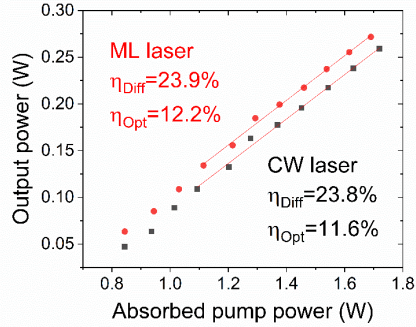

Fig. 6. Output parameters of the oscillator in different operation modes.

Maximum average output power in the ML regime is slightly higher than in CW – 273 mW versus 259 mW with optical-to-optical efficiencies of 12.2% and 11.6% respectively, which isdue to the soft-aperture based self-amplitude modulation. Maximum pulse energy at the CPO output was about 22 nJ. In addition to power/energy characteristics of the CPO we also measured pulse duration (blue curve) and spectral width (green curve in Figure 7).

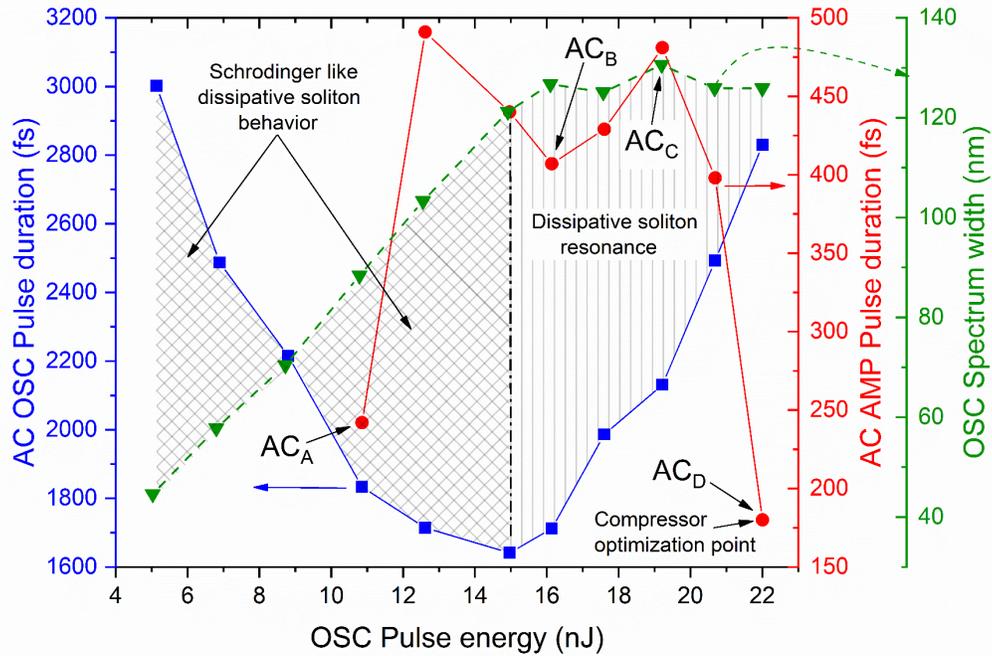

Fig. 7. Autocorrelation traces' durations and oscillator spectral width in dependence of oscillator output pulse energy. OSC: oscillator output. AMP: amplifier-compressor output.

As one can see from the data, the CPO pulse duration (uncompressed, blue curve on Figure 7) – spectral width (green curve at Figure 7) behavior can be subdivided into two zones. First zone with pulse energies from 5 to 15 nJ, where pulse duration shortens and spectral width increases, and the second zone with pulse energies from 15 to 22 nJ, where spectral width stays nearly the same, but pulse duration approximately linearly increases. The first zone behavior has some similarities to the Schrödinger soliton, but the second zone is more interesting for our power scaling purposes. In this case the DS spectral width is clamped in a certain range, defined by the CPO design parameters, so the pulse must increase its duration in order to accommodate the energy growth into the CPO area theorem. Such behavior is called "*Dissipative soliton resonance*" (DSR) in the literature [24] and is achieved by increasing the pulse phase modulation (chirp) thus making the pulse longer. In Fig. 7, the red line shows the amplified pulse duration after the compressor, which was optimized for the highest energy. As expected, the autocorrelation trace durations indicate overcompensation at lower energies, where the chirp is lower.

The autocorrelation traces and pulse spectra for three characteristic points (A – before DS resonance, B – transition period and C and D – DS resonance, on the Figure 7) are shown in Figs. 8 and 9.

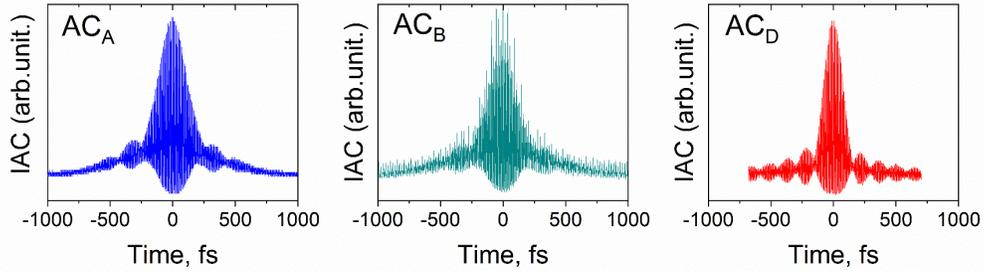

Fig. 8. Interferometric autocorrelation traces of the amplified pulses for specific points on the CPO regime diagram (see Fig. 7).

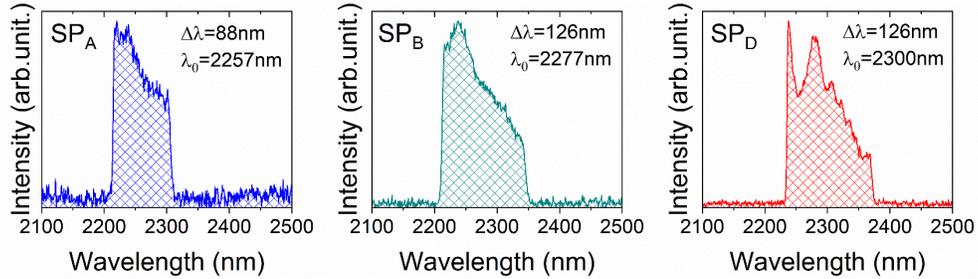

Fig. 9. Amplified pulse spectra for specific points on the CPO regime diagram (see Fig. 7).

A short comment on the spectral width estimation technique. DSs at the CPO output have specific shape of the spectrum namely, a spectrum with sharply cut edges. In addition, the incompletely compensated TOD in our laser leads to a slope in the upper part of the spectrum, which introduces additional ambiguity in the case of using the width at half maximum criterion. We therefore used the following width definition: we calculated the first derivative of the spectral curve and estimated the distance (spectrum width) between the peaks of the derivative (see Figure 10).

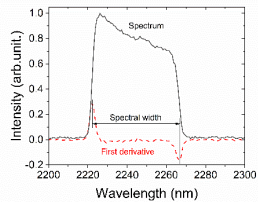

Fig. 10. Illustration of spectral width estimation technique.

Input pulse energy was 20 nJ resulting in output pulse energy of 62 nJ under amplifier incident pump power of about 12.7 W result in 6% optical-to-optical efficiency (energy gain ratio 3.1). Taking into account gain and pump absorption saturation, we estimated the gain to reach ~3.8 for this pumping power. The difference the with experiment is probably due to uncertainty in determining the sizes of the pump beam and CPO output beam waists in the amplifier gain medium. The small signal pump absorption in the amplifier active medium was about 80%, dropping to just above 40% at 12.7 W due to the absorption saturation. The compressor transmission of about 80% reduces the output pulse energy to 50 nJ that correspond to 278 kW power.

It should be noted, that using the CPO seed effectively turns this setup to a chirped-pulse amplifier scheme with strongly reduced nonlinearity. The spectra of the amplified and seed pulses are nearly identical (as opposed to soliton amplification [31]) with the same width and central wavelength, for example at maximum output power of the system ($SP_D$ point on Figure 7) spectral width was about 126 nm at a central wavelength of 2300 nm, indicating negligible spectral phase distortion and suggesting further energy scalability of this scheme.

## 4. DS scalability and compressibility

The advantage of CPO as a source for further pulse amplification is its energy scalability provided by a significant DS chirp [17]. It is well-known [36, 37] that a DS developing in the normal GDD regime ($\beta_2 > 0$) can be described as a soliton-like solution of the complex cubic-quintic non-linear Ginzburg-Landau equation [38]:

$$\frac{\partial a(z,t)}{\partial z} = \left\{ i \left( \beta_2 \frac{\partial^2}{\partial t^2} - \gamma |a(z,t)|^2 \right) + \left[ -\sigma + \alpha \frac{\partial^2}{\partial t^2} + \kappa(1 - \zeta |a(z,t)|^2) |a(z,t)|^2 \right] \right\} a(z,t). \quad (1)$$

Here, $\beta_2$ is a GDD coefficient, $\alpha$ is a squared inverse bandwidth of a spectral filter (e.g., a gain bandwidth), $\gamma$ is a self-phase modulation (SPM) coefficient, $\sigma$ is a saturated net-loss coefficient, $\kappa$, and $\zeta$ are the coefficients describing a saturable nonlinear gain due to soft-aperture mode-locking [39].

*Adiabatic theory* of a strongly-chirped DS [26, 36] allowed a unified description of DS by mapping a parametric system space to the so-called *master diagram* (Figure 11) spanned by two dimensionless coordinates: $c = 2\alpha\gamma/\beta_2\kappa$ and $E^* = E \times \left( \kappa \sqrt{\zeta/\beta_2 \gamma} \right)$, where the first parameter $c$ describes a relative contribution of spectral dissipation, GDD, self-phase and self-amplitude modulations. The second coordinate $E^*$ is a dimensionless DS energy $E$ [36]:

$$E = \frac{6\gamma}{\zeta \kappa} \Xi^{-1} \arctan \frac{\Delta}{\Xi}. \quad (2)$$

The complex spectral amplitude $\varepsilon(\omega)$ in the absence of higher-order GDD has the following form [40]:

$$\varepsilon(\omega) = \sqrt{\frac{6\pi \gamma}{\kappa \zeta}} \frac{\exp\left[\frac{3i\gamma^2\omega^2}{2\kappa\zeta\beta_2(\Xi^2+\omega^2)(\Delta^2-\omega^2)}\right] \text{Heaviside}(\Delta^2-\omega^2)}{\sqrt{i(\Xi^2+\omega^2)}}. \quad (3)$$

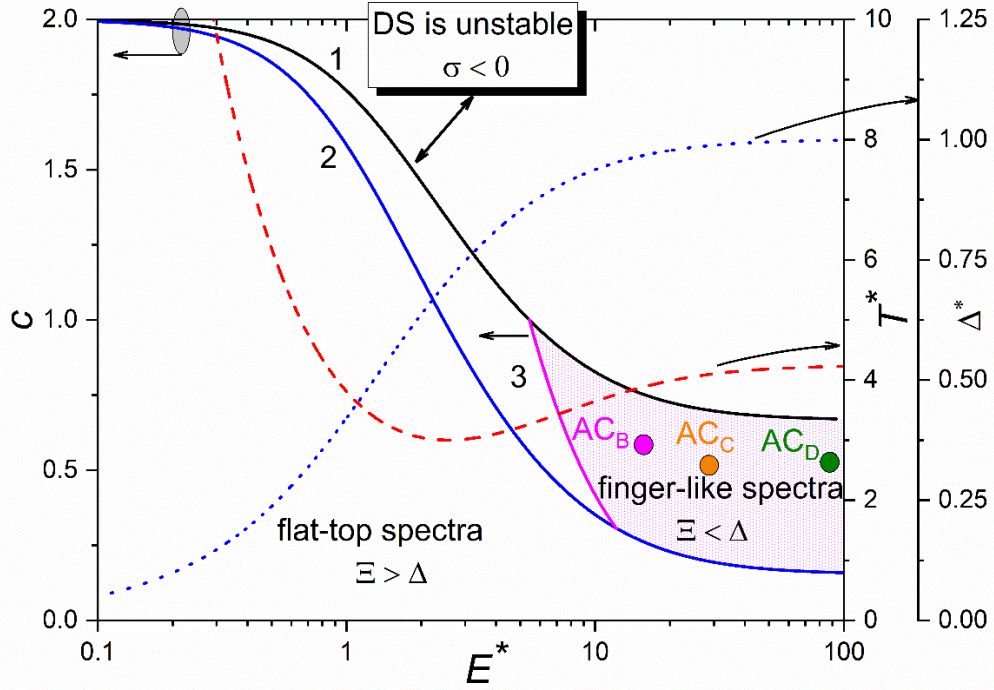

Fig. 11. Master diagram representing DS stability threshold (black curve 1), bottom boundary of RES (blue curve 2), spectral half-width $\Delta^*$ (blue dotted curve), and DS width $T^*$ (red dashed curve) in dependence on the normalized energy $E^*$. Magenta curve 3 confines a region of energy-scalable DS with finger-like spectra. The points correspond to the experimental ones in Fig. 7. Normalizations are: $E^* = E \times \left(\kappa\sqrt{\zeta/\beta_2 \gamma}\right)$, $T^* = T \times \left(\kappa/\sqrt{\gamma \zeta \beta_2}\right)$, and $\Delta^* = \Delta \times \sqrt{\beta_2 \zeta/\gamma}$.

The parameter $\Delta = \sqrt{\gamma P_0/\beta_2}$, where $P_0$ is a DS peak power, defines the spectrum half-width, i.e., the cut-off frequency. Formally, such a cut-off can be explained by a resonance between DS and background radiation with the wave-numbers $\gamma P_0$ and $\beta_2 \Delta^2$, respectively [29, 41].
A Lorentz-like central profile ("finger") of the spectrum is defined by a "width" $\Xi = \sqrt{\gamma\left((1+c)/\zeta - \frac{5P_0}{3}\right)/\beta_2}$. The characteristic spectral power profile from (3) is

$$p(\omega) = |\varepsilon(\omega)|^2 = \frac{6\pi\gamma \, \text{Heaviside}(\Delta^2 - \omega^2)}{\zeta\kappa(\Xi^2 + \omega^2)} \qquad (4)$$

and the corresponding spectral chirp is

$$Q(\omega) = \frac{1}{2}\frac{d^2\phi(\omega)}{d\omega^2} = \frac{3\gamma^2}{2\beta_2\kappa\zeta}[(\Delta^2 - \omega^2)(\Xi^2 + \omega^2)]^{-1}. \qquad (5)$$

Fig. 11 demonstrates that the DS energy is scalable along curve 1 (black) $\sigma = 0$, defining the DS stability. In the same way as for a Schrödinger soliton, the negativity of net-gain $\sigma$ is a necessary condition of DS stability against the background noise. In a steady-state, this parameter can be described as $\sigma \approx \delta(E/E_{cw} - 1)$, where $E_{cw}$ is an energy of CW-radiation, i.e., a mean power multiplied by a cavity period in the CW-regime, and $\delta = l^2/g_0$ ($l$ is a net-loss coefficient, $g_0 = l\left(E_{cw}/E_s + 1\right)$ is a small-signal gain, and $E_s$ is a gain saturation energy) [26]. The implicit expression for the stability threshold $\sigma = 0$ presented by the black solid curve in Fig. 11 follows from Eq. (2), when $\Delta$ and $\Xi$ are taken for a limiting value $P_0 = (3/2)(1 - c/2)\zeta^{-1}$.

One has to note that the notion of energy scalability, which is equivalent to DSR [25], means that the DS energy can be scaled only by the scaling of pump, beam area, and cavity period without a change of other parameters defining $c$ (i.e., without a change of GDD first of all). From Eq. (2), such scalability (i.e., $\lim_{\Xi \to 0} E = \infty$) exists when $\Xi \to 0$, i.e., when $c = \frac{2}{3}$ (high-energy asymptotics of black curve 1 in Fig. 11 corresponding to the stability threshold $\sigma = 0$). We would summarize the DSR parameters based the adiabatic theory:

$$E \to \infty \Leftrightarrow \begin{cases} P_0 \to \zeta^{-1}, \\ \Delta \to \sqrt{\gamma/\beta_2\zeta}, \\ \Xi \to 0, \\ c = 2/3, \\ b = 0. \end{cases} \tag{6}$$

The condition of $b = 0$ ($b = \zeta\sigma/\kappa$) physically means that the upper boundary of the *region of energy scalability* (RES) corresponds to the threshold of the DS stability, i.e., the condition of $\sigma = 0$, when a saturated gain equals net-loss in a laser. Obviously, that real $\sigma$ should be positive but close to zero. In this case, RES is bounded from below by $b \in [0, \approx 0.21]$ and $c \in [2, \approx 0.15]$ (the solid blue curve 2 in Fig. 11):

$$\max E \, (b > 0) \Leftrightarrow \begin{cases} E = \frac{6\sqrt{2\gamma\beta}}{\kappa\sqrt{\zeta}} \frac{\arctan\left(\frac{\sqrt{3}\sqrt[4]{b}}{\sqrt{6-13\sqrt{b}}}\right)}{\sqrt{(6-13\sqrt{b})}}, \\ b \in \left[0, \frac{36}{169}\right], \\ c = 2 - 4\sqrt{b}, \\ P_0 = \frac{3\sqrt{b}}{2}\zeta^{-1}, \\ \Delta^2 = \frac{3\sqrt{b}\gamma}{2\beta\zeta}, \\ \Xi^2 = \frac{\gamma}{2\zeta\beta}(6 - 13\sqrt{b}). \end{cases} \tag{7}$$

Both curves have the asymptotic $E \to \infty$ for $c = \frac{2}{3}, 2(1 - 2\sqrt{36/169})$ corresponding upper and bottom boarders confining RES.

The energy-scaling is provided by the DS broadening (see the dashed red curve in Fig. 11 corresponding to DSR for $b = 0$ and compare with the solid blue one in Fig.7) with an asymptotically constant spectral half-width $\Delta \in [\approx 0.83, 1] \times \sqrt{\gamma/\beta_2\zeta}$ (compare Figs. 7 (dashed green curve) and 11 (dotted blue curve)), $\Xi \to 0$, and the confined peak power $P_0 \in [\approx 0.7, 1] \times \zeta^{-1}$.

Figs. 7, 11, and Eqs. (6) demonstrate the hallmarks of a transition to DSR within RES: 1) change of pulse squeezing to its broadening with growing energy, 2) constant spectral width $\Delta$, and 3) appearance of a visible spike at the spectrum center. The *finger-like* spectrum ($\Xi < \Delta$) (see Eq. (5), Fig. 12 and [26, 36, 42]) is clearly visible in Figure 9, $SP_C$. The first hallmark is crucial for DS energy harvesting: its peak power is fixed (Eq. (6)), but the energy grows by the pulse stretching $\propto 1/\Xi$ (Figs. 7, 11) due to a chirp scaling [26]: $\psi \cong T\Delta$

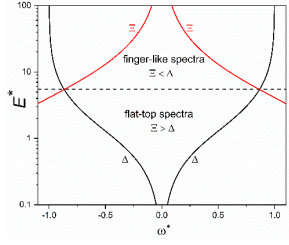

Fig. 12. Behavior of spectral half-width $\Delta^*$ ("cut-off frequency", black curves) and "finger half-width" $\Xi^*$ ("chemical potential", red curves) in dependence on the normalized energy $E^*$. Normalizations are: $E^* = E \times \left(\kappa\sqrt{\zeta/\beta_2\gamma}\right)$, and $\omega^* = \omega \times \sqrt{\beta_2\zeta/\gamma}$.

The latter factor allows energy re-distribution inside a DS, preserving its integrity. Using the approximating approach for a DS profile (Figure 13)

$$a(t) = a_0 \operatorname{sech}\left(\frac{t}{T}\right) \exp[i(\phi(z) + \Omega t + \theta \tanh\left(\frac{t}{T}\right) + \psi \log \operatorname{sech}\left(\frac{t}{T}\right) + \chi \log^2 \operatorname{sech}\left(\frac{t}{T}\right))] \quad (8)$$

($\phi$ is a phase, $\Omega$ is a frequency shift from a gain band centrum, $\psi$ is a chirp, $\theta$ and $\chi$ are the phase distortions cause by TOD and FOD, respectively) allows defining the energy flow $j(t')$ inside the soliton ($t' = t/T$, $T$ is a DS width) [43, 44]:

$$j(t') \equiv \frac{i}{2}(a\,\partial_{t'}a^* - a^*\partial_{t'}a) =$$

$$= \frac{a_0^2}{2}\operatorname{sech}^4(t')\left[\left(\psi + 2\chi \log \operatorname{sech}\left(\frac{t'}{T}\right)\right)\sinh(2t') - 2\theta\right] \quad (9)$$

Figure 14 demonstrates such a stabilizing flow, which increases with $\psi$.

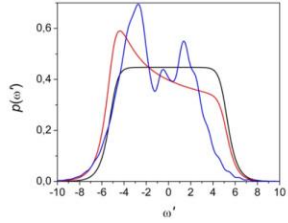

Fig. 13. Dimensionless spectral profiles $p(\omega)$ for $\psi = 5$, $\theta = \chi = \Omega = 0$ (black curve), $\psi = 5$, $\theta = 2$, $\chi = \Omega = 0$ (red curve) and $\psi = 5$, $\theta = 2$, $\chi = 1$, $\Omega = 0$ in Eq. (8).

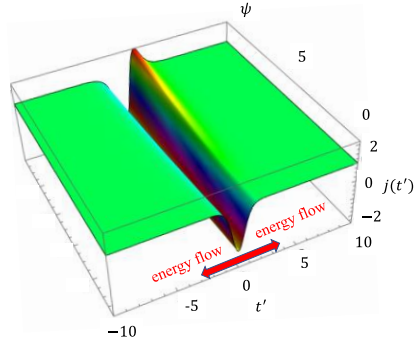

Fig. 14. Dimensionless energy flux $j(t')$ in dependence on the chirp $\psi$, $\theta = \chi = 0$ in Eq. (8).

As Figure 13 demonstrates, the observed asymmetry of spectra (Figs. 9, 10) is a direct result of the TOD (Figure 4). This factor could be an obstacle to the energy scaling and needs elimination by a fine GDD control in the CPO. An additional factor affecting the DS spectrum is FOD causing the formation of an M-like spectral shape (Figure 13) [45].

The next practically important phenomenon demonstrated by the master diagram is the transition to an energy-scalable regime at $c = \sqrt{9-48b}/3$ when $\Delta = \Xi$ (Fig. 12) so that:

$$
\begin{aligned}
c &= \frac{\sqrt{9-48b}}{3}, \\
E &= \frac{3\sqrt{2}\pi\sqrt{\gamma}}{\kappa\sqrt{\zeta\beta(3+\sqrt{c})}}, \\
\Delta &= \Xi = \frac{\sqrt{6\gamma(1+c)}}{4\sqrt{\zeta\beta}}, \\
P_0 &= \frac{3}{8\zeta}(1+c)
\end{aligned}
\tag{10}
$$

within the interval $b \in [0, \approx 0.17]$ (magenta curve 3 in Figure 11). That corresponds to a maximally flat spectral chirp $Q(\omega)$ around the DS spectral center (Eq. (5)), which means the maximal DS compressibility (or *maximum compression fidelity*) by a subsequent chirp compensation in a compressor [46]. Generally speaking, the magenta curve 3 in Fig. 11 represents the energy border of domain of maximal DS energy scalability with potentially the best compressibility.

The compensating anomalous dispersion along the fidelity curve in Fig. 11 can be easily estimated from Eq. (3). Since the spectral chirp is maximally flat around $\omega = 0$, one could adjust the compensating anomalous dispersion $\delta$ to the central frequency so that Eq. (3) can be rewritten for the dimensionless values as:

$$
\varepsilon^*(\omega) = \sqrt{6\pi} \frac{\exp\left[\frac{3id\omega^{*2}}{2(\Xi^{*2}+\omega^{*2})(\Delta^{*2}-\omega^{*2})} - i\delta\omega^{*2} - i\pi/4\right] \text{Heaviside}\left(\Delta^{*2}-\omega^{*2}\right)}{\sqrt{\Xi^{*2}+\omega^{*2}}},
\tag{11}
$$

Where $d = \gamma/\kappa$, and the frequencies and spectral amplitude are normalized to $\sqrt{\kappa\beta_2/\gamma}$ and $\sqrt{\zeta/\beta_2}$, respectively. Along the fidelity curve, the chirp flatness is provided by $(\Xi^{*2}+\omega^{*2})(\Delta^{*2}-\omega^{*2}) = \Delta^4 - \omega^4$ [36]. From Eqs. (10,11), it is easily to estimate the dimensionless compensating dispersion:

$$
\delta = \frac{32d}{3(1+c)^2} = \frac{96d}{(3+\sqrt{9-48b})}.
\tag{12}
$$

It is evident from (7,11,12), that the compensating dispersion is defined by chirp $\propto P_0 \propto \zeta^{-1}$ and grows with it.

Our message is that CPO could inject an energetical and high-fidelity seed with the phase characteristics providing a multi-stage amplification with the final compression down to a minimal pulse width and maximal peak power without the energy loss into satellites.

The following equations [36]

$$
\begin{aligned}
\sigma &= \vartheta\left(\frac{E}{E_{cw}} - 1\right), \\
P_0 &= \frac{\beta\Delta^2}{\gamma}, \\
\beta\Xi^2 + \frac{5}{3}P_0\gamma &= \frac{\gamma}{\zeta}(1+c), \\
\zeta P_0 &= \frac{3}{4}\left(1 - \frac{c}{2} + \sqrt{\left(1-\frac{c}{2}\right)^2 - 4\zeta\sigma/\kappa}\right).
\end{aligned}
\tag{13}
$$

plus Eq. (2) provide us with a guide to approximate estimation of unmeasurable parameters of the self-amplitude modulation $\kappa$ and $\zeta$ in Eq. (1). There are the experimental data for the average power in the CW and mode-locking regimes, which, in the combination of data for the net-losses $l$ and the cavity period, give $E_{cw}$ and $E$ in the first equation of (13). Then, the "stiffness" parameter is $\vartheta = l^2/g_0$ for the small signal gain $g_0 = l(1 + E_{cw}/E_{sat})$, where $E_{sat}$ is the gain saturation energy [26]. As a result, one has the $\sigma$-value (Eq. (13) and Table 1).

The net-GDD and SPM parameters, as well as the DS spectral half-width $\Delta$ are known (Fig. 4 and Table 1), which allows obtaining the value of the DS peak power $P_0$ from the second equation (13). The remaining two equations (13) plus Eq. (2) give unknown $\kappa, \zeta$, and $c$ for the experimentally measured $\Xi$ (the last is obtained as a Lorenzian fitting of a "finger" on the experimental DS spectra). The result can be fitted in the master diagram ($AC_B$, $AC_C$, and $AC_D$ points in Fig. 11). The obtained estimations for $\kappa$ and $\zeta$ are close to those for a soft-aperture systems [39]. The $AC_B$ point is very close to the border of the energy scalability region, where the spectrum width tends to asymptote (Figs. 7,11). On the left side of this border, the Lorentzian peak becomes unrecognizable so that the pulse width could play the role of the third equation in (13) [36]:

$$T = \frac{3\gamma^2}{\zeta\beta\kappa\Delta(\Delta^2+\Omega^2)}. \tag{14}$$

Table 1. Experimental and theoretical parameters relevant to the master diagrams in Figs. 7,11.

Experimental data

| | |
|---|---|
| central wavelength $\lambda$, cm | $2.3 \cdot 10^{-4}$ |
| gain cross-section $\sigma$, cm | $1.38 \cdot 10^{-18}$ |
| gain relaxation time $T_r$, $\mu$s | 4.9 |
| cavity period $T_{cav}$, ns | 81 |
| mode area $S$, $\mu m^2$ | 2513 |
| gain saturation energy $E_s = h\nu \frac{T_{cav}S}{T_r \sigma}$, nJ | 186 |
| out-put loss $l$, % | 14 |
| active crystal length $z$, cm | 0.26 |
| nonlinear refraction coefficient $n_2$, cm$^2$W$^{-1}$ | $9 \cdot 10^{-15}$ |
| self-phase modulation coefficient $\gamma$, MW$^{-1}$ | 5.1 |
| net-GDD coefficient $\beta_2$, fs$^2$ | 440 |
| pulse spectrum width for the points $AC_C$ a $AC_D$, nm | 126 |

Theoretical estimations

| | | | |
|---|---|---|---|
| small signal gain $g_0$ ($AC_B$, $AC_C$, $AC_D$) | 0.12 | 0.129 | 0.138 |
| "stiffness" parameter $\vartheta$ | 0.055 | 0.048 | 0.046 |
| intracavity CW energy $E_{cw}$, nJ | 98.5 | 121 | 143 |
| intracavity pulse energy $E$, nJ | 110.5 | 132 | 151 |
| saturated net-loss coefficient $\sigma$ | 0.0066 | 0.0046 | 0.0023 |
| dimensionless energy $E^*$ | 15.6 | 27.9 | 87.4 |
| pulse spectrum half-width $\Delta$, THz | | 22.4 | |
| Lorenzian width $\Xi$, THz | 13.4 | 8 | 5.34 |
| self-amplitude modulation parameter $\kappa$, MW$^{-1}$ | 2.3 | 3.5 | 4.7 |
| self-amplitude modulation saturation parameter $\zeta$, MW$^{-1}$ | 29 | 32 | 34 |
| chirp parameter $\psi$ | 38 | 48 | 64 |

## 5. Discussion

A qualitative comparison of experimental and theoretical results provides a roadmap for a hybrid CPO-CPA pulse energy scaling. As the master diagram (Figure 11) demonstrates, the operating parameters of a CPO have several distinct areas separated by the stability and maximum fidelity curves. Excluding the area of instability, we can focus on the two regions on the master diagram divided by the maximal pulse fidelity curve.

The characteristic feature of this division is the behavior of the pulse spectrum, which changes its shape from a flat-top to a finger-like one when passing through the maximum fidelity curve in the direction of increasing pulse energy. Such behavior is clearly seen during our experiment- spectral shape changes from flat-top (Figure 9, SP$_A$) to finger-like (Figure 9, SP$_C$ and SP$_D$), which confirms the CPO operation near the maximum fidelity curve. Then, crossing a fidelity border accompanies a change of the DS width behavior with the energy growth- from decreasing to rising (Figs. 7, 11b) with the spectral width finally reaching the asymptotic. Such qualitative behavior of the sound and measurable CPO characteristics provides a path to CPO design and optimization.

These hallmarks follow the idea of DS energy scaling with maximum fidelity (compressibility) and using such extremely broad and smooth gain bandwidth media as $Cr^{2+}$:ZnS without a significant spectral phase degradation in CPA. That could be a road to developing the high-power ultrafast laser system with ultrashort pulse durations working at high pulse repetition rates.

One must remark on the main obstacles to realizing the perfect DS power/energy harvesting in a CPO-CPA system. The first is a destructive contribution of higher-order dispersions (TOD, FOD, etc.). They result in the DS spectrum distortion and squeezing and may cause a strong destabilization, including chaotization and even DS fission [30, 45, 47]. Especially, the TOD is maximally destructive in this way that requires applying the elaborated dispersion compensation techniques.

There also exists the risk of fidelity degradation with advancing into the domain of perfect energy scalability, which needs an additional study. The issue could be illustrated by a close analogy between DS and turbulence [48]. The energy harvesting results from a soliton width growth $\Lambda_l \propto 1/\Xi$. The last parameter $\Lambda_l$ plays a role of a "long-range" correlation scale [49]. Simultaneously, the "cut-off" frequency $\Delta$ defines a "short-range" correlation scale $\Lambda_s \propto 1/\Delta$. Under this angle of view, it is the coincidence of these scales $\Lambda_l = \Lambda_s$ that provides maximal fidelity. However, the growing discrepancy of these correlation scales with the $\Xi$-decrease can lead to a DS "decoupling", that is to the emergence of "soliton gas" at a "distance" of $\Lambda_s$ [50, 51] trapped by a collective "potential" with a characteristic width $\propto \Lambda_l$. Such partially coherent DS becomes more sensitive to perturbations, which are clearly visible in the spectrum (see Fig. 9, SP$_D$ and [52]).

Finally, from a thermodynamical point of view, the expression (4) is analogous to the Rayleigh-Jeans distribution [53, 54] with a negative "chemical potential" $\mu = \Xi^2$ and a "temperature" $T = 6\pi\gamma/\zeta\kappa$ [48]. The DS energy growth can be treated as a result of the particle number $N$ (mass) increase by analogy with the Bose-Einstein condensation from a "basin" [55]. The negativity of chemical potential preserves the entropy $S$ in such a process when a temperature is constant and the Gibbs free energy $dF = dU - TdS - SdT$ decreases, in agreement with the fundamental thermodynamic relation $dU = TdS + \mu dN$, where $U$ is a total energy of a system. However, $\mu$ tends to zero in parallel with such a process (Eq. (6) for $c = 2/3$). This "wave condensation" counteracts the further particle number growth in parallel with the $F$ minimization. By analogy with a Bose-Einstein condensate, it means that a further condensation (i.e., DS energy growth) becomes impossible [56], and DS tends to the internal decoupling ("fission") due to $\Lambda_s/\Lambda_l \to 0$. However, the detailed analysis of the transit to turbulence awaits further study.

## 6. Conclusion

Summing up, in this paper we suggest theoretically and show experimentally feasibility of a novel technique to generate high peak power laser pulses and further energy scale them in a mid-IR chirped pulse oscillator-amplifier (CPO-CPA) system, without sacrificing neither the pulse duration nor energy. The concept allows avoiding the destructive nonlinearities in the final stages of amplifiers and compressor elements by using a chirped high-fidelity pulse from a CPO as a seed for a single-pass CPA. Such scheme eliminates the necessity of stretchers and expensive and troublesome electrooptical instrumentation, being suitable for pulse generation and amplification systems operating at high pulse repetition rates. A qualitative comparison of experimental and theoretical results provides an algorithmically realizable roadmap for a hybrid CPO-CPA pulse energy scaling while not having to sacrifice the output pulse duration by preserving the output spectrum's high-fidelity compression. Experimentally, we demonstrate a $Cr^{2+}$:ZnS laser system that provides 180 fs pulses with 50 nJ pulse energy at 12.345 MHz pulse repetition frequency. With higher-chirp CPO and more amplification stages the pulse energy can reach µJ levels.

The proposed methodology promises to open up a way to compact table-top laser instruments that combine high repetition rate and high peak power in the mid-IR, enabling productivity increase in applications and novel scientific approaches, that are yet inaccessible neither in university laboratories nor in the industry.

**Funding.** The work is supported by the Norwegian Research Council projects #303347 (UNLOCK), #326503 (MIR), ATLA Lasers AS.

**Acknowledgements.** The work of AR, MD, VLK and ITS was supported by NFR projects #303347 (UNLOCK), #326503 (MIR), and by ATLA Lasers AS.

**Disclosures.** ITS: ATLA Lasers AS (I,S) , ES: ATLA Lasers AS (I).